\documentclass[times]{elsarticle}
\usepackage[utf8]{inputenc}

\usepackage[margin = 1in]{geometry}
\usepackage{booktabs}
\usepackage{amsmath}
\usepackage{hyperref}
\usepackage{amssymb}
\hypersetup{colorlinks=true, linkcolor=blue, citecolor=blue, urlcolor=blue}
\usepackage[]{natbib}
\usepackage{soul}

\newcommand{\bs}[1]{\boldsymbol{#1}}
\newcommand{\probz}{\mu_{\bs{z}}(\bs{z})}
\newcommand{\probx}{\mu_{\bs{x}}(\bs{x})}

\newcommand{\aprobx}{q_{\bs{x}}(\bs{x})}
\newcommand{\kb}{k_B}
\newcommand{\mref}[1]{(\ref{#1})}
\date{\today}

\begin{document}

\begin{frontmatter}
    \title{Free energy calculation of crystalline solids using normalizing flow}

    \author[label1]{ Rasool Ahmad \corref{cor1}}
    \ead{rasool@stanford.edu}
    \author[label1]{Wei Cai}
    \ead{ caiwei@stanford.edu }
    \address[label1]{Micro and Nano Mechanics Group, Department of Mechanical Engineering,
        Stanford University, CA 94305, USA}
    \cortext[cor1]{corresponding author}

    \begin{abstract}
        Taking advantage of the advances in generative deep learning, particularly normalizing flow, a framework, called Boltzmann Generator, has recently been proposed for the purpose of generating equilibrium atomic configurations from the canonical ensemble and determining the associated free energy. In this work, we revisit Boltzmann Generator to motivate the construction of the loss function from the statistical mechanical point of view, and to cast the training of the neural networks in purely unsupervised manner that requires no samples of the atomic configurations from the equilibrium ensemble. We further show that the normalizing flow framework furnishes a reference thermodynamic system, very close to the real thermodynamic system under consideration, that is suitable for the well-established free energy perturbation methods to determine accurate free energy of solids. We then apply the normalizing flow to two problems: temperature-dependent Gibbs free energy of perfect crystal and formation free energy of monovacancy defect in a model system of diamond cubic Si. The results obtained from the normalizing flow are shown to be in good agreement with that obtained from independent well-established free energy methods.
        %We then apply the framework to compute the Gibbs free energy of diamond-cubic crystal of Si, modeled using the Stillinger–Weber potential, at finite temperatures, and demonstrate a good agreement with the results that are obtained from independent free energy methods.
    \end{abstract}

\end{frontmatter}

\section{Introduction}

% this paragraph describes importance of free energy and difficulty in computing it from atomistic simulations, establishing the motivation of this work.
Free energy is an important thermodynamic quantity that characterizes the behavior of materials at finite temperature. For instance, free energy determines the relative stability of different phases of materials~\cite{Gillan2006, Broughton1987}, population of point defects~\cite{Cai2016}, and rate at which thermally activated processes occur in solids~\cite{Vineyard1957,Calliard2003,Bulatov2006a}. Computation of free energy from atomistic simulations such as molecular dynamics (MD) and Monte Carlo (MC) simulations, however, is challenging due mainly to the inability to express free energy as an ensemble average, and difficulties involved in the computation of the partition function, the normalizing constant of the probability distribution function of atomic microstate~\cite{Frenkel2001, Tuckerman2011}. The importance of free energy in understanding the finite temperature behavior has prompted the development of several techniques based on  equilibrium and non-equilibrium statistical mechanics which are summarized below.

% brief overview of various practical free energy method and emphasizes the need of a reference system in free energy methods.
Most of the free energy methods start with the choice of a reference system of which free energy is
exactly known \cite{Chipot2007}. Subsequently, a switching path is designed between the reference system and the system of interest. In one class of methods, the switching between the two systems is carried out along the path such that the system always remains in thermodynamic equilibrium. The free energy energy difference between the the two systems is then simply the work done along the switching path, and is determined by performing thermodynamic integration (TI) with the help a chain of finite temperature atomistic simulations \cite{Kirkwood1935, Frenkel2001}. Another class of free energy methods make use of time dependent non-equilibrium trajectories between the two systems and relates the work done during the slow switching to the free energy difference between the two systems~\cite{Watanabe1990, DeKoning1997, Freitas2016,Jarzynski1997}.  In both equilibrium and non-equilibrium methods, the choice of reference system critically influences the computational efficiency of the method~\cite{Tsao1994, DeKoning1996}. The closer the two systems involved in the switching are in the phase space, the more accurate is the computed free energy.

% use of generative deep learning methods in sampling atomic configuration from the canonical (NVT) ensemble.
The rapid advances in the field of machine learning are being utilized in building powerful tools to  analyze, augment and accelerate the atomistic/molecular simulations on several fronts~\cite{Bartok2017a, Noe2020, Ceriotti2021}. Since finite temperature atomic simulations fundamentally aim to generate the atomic configurations from the underlying equilibrium probability distribution function, we focus our attention to generative machine learning. Generative deep learning has offered powerful techniques to learn the underlying probability distribution of objects such as images, and subsequently generate samples from the learned distribution. Generative Adversarial Networks (GANs) \citep{Goodfellow2014} have gained prominence in generative modeling due to its remarkable ability to generate life-like fake images of imaginary human beings and other objects. GANs, however, does not provide access to explicit probability density of objects being generated and thus limiting the application of GANs in situations where probability density of generated samples are required. Normalizing flow~\cite{Rezende2015}, on the other hand, is another rapidly growing generative modeling framework which applies a parametric bijective transformation to a simple base distribution, for example Gaussian distribution, in order to variationally approach the target probability distribution. The normalizing flow framework is quite flexible in that it seeks to closely approximate the probability distribution function of a given set of data, generate new samples from the learned probability distribution function, and provide an associated probability density to each generated sample \citep{Kobyzev2020, Papamakarios2019}. In spite of some limitations regarding expressivity, normalizing flow can have a significant impact on the way atomic configurations are generated during atomistic simulations, and free energy is determined.

% discussion on the use of normalizing flow to physics problems.
For instance, No\'e et al \cite{Noe2019a} introduce the Boltzmann generator, an algorithm based on normalizing flow, that is capable of drawing equilibrium atomic microstates and compute the free energy difference associated with different configurations of molecules and condensed matter systems under the canonical ($NVT$) conditions. Flow based models have further found numerous applications in solving physics problems: Albergo et al \cite{Albergo2019a} generate  samples from a probability distribution in field theories; Nicoli et al \cite{Nicoli2020} employ normalizing flow for asymptotically unbiased estimation of physical observables; Muller et al \cite{Muller2019} compute Monte Carlo integration utilizing normalizing flow; Xie et al \cite{Xie2021} study interacting fermions at finite temperature; Li and Wang \cite{Li2018a} employ normalizing flow to construct a hierachical renormalization group approach to identify independent collective variables. Application of normalized flow based algorithms to computing absolute free energy of real solid systems and systematic comparison with the existing methods has, however, to the authors' best knowledge, not been fully explored.

% primary goal and a summary of the findings of the present work (compare the efficiency and computational cost of the two methods.)
The present study, building on the work of No\'e et al \cite{Noe2019a}, primarily focuses on to demonstrating that normalizing flow can be employed to compute absolute free energy of solids with the same level of accuracy as that obtained from widely used free energy methods. In particular, we use diamond-cubic Si crystal as our model solid system to compute its Gibbs free energy at temperatures ranging from 1600 to 2000 K and compare it with the adiabatic switching method. We, additionally, consider the formation free energy of monovacancy in diamond-cubic Si at temperatures ranging from 0 to 300 K. The normalizing flow values for both cases are shown to agree quite well with other well-established free energy methods.

In this work, normalizing flow is used to learn the distribution of atomic displacement from a fixed atomic structure that is essentially the metastable structure corresponding to the local minima of the potential energy surface. Normalizing flow is shown to provide a good reference system to be used in well-established free energy perturbation methods that remove any bias originating from the practical limitation on the available computational resources. To mitigate the difficulties arising from the uncontrolled motion of the center of mass of the crystal due to the translation invariance of the potential energy function, we introduce an additional harmonic potential energy term for the center of mass motion in the potential energy function. The final free energy is then obtained by subtracting the contribution of the center of mass motion. Another interesting feature of normalizing flow as employed in this work is that the training of normalizing flow proceeds in a purely unsupervised manner in the sense that it does not require a set of atomic structures drawn from the equilibrium ensemble as examples to learn from. Thus, in the terminology of No\'e et al~\cite{Noe2019a}, the training is performed only by energy, and not by example. The data for training are generated from the base distribution which are potentially limitless in principle, and mitigate the concern of overfitting.

% structure of the manuscript.
The rest of the manuscript is arranged in the following manner. Section \ref{sec_free_eng_pert} briefly describes free energy perturbation method that inspires various practical free energy methods. Next, the basic working of normalizing flow and the use of variational free energy as the loss function for training are presented in Section \ref{sec_norm_flow}. Section \ref{sec_reweighting} provides details on how the transformed distribution obtained from normalizing flow can be used as the reference system in free energy perturbation methods. Section \ref{sec_application_si} presents an application of normalizing flow to diamond cubic Si with information about computational methods and relevant parameters. Finally, Section \ref{sec_conc_disc} recapitulates important points and discusses various potential extensions, modifications, and future applications of normalizing flow.

\section{Free energy perturbation method}
\label{sec_free_eng_pert}

% introduces Boltzmann distribution as the probability distribution of atomic configuration in the canonical ensemble.
In this section, we present the basic concepts involved in free energy perturbation method following Zwanzig~\cite{Zwanzig1954}. We mainly focus our discussion on the canonical $(NVT)$ ensemble in which the number of atoms $N$, volume $V$, and temperature $T$ are kept fixed. Helmholtz free energy is the relevant free energy in the canonical ensemble, and hereafter the free energy should be understood as referring to the Helmholtz free energy. The canonical ensemble is characterized by the Boltzmann distribution of atomic microstates/configuration, i.e.  the joint probability distribution of $3N$ dimensional vector of atomic positions $\bs{x}$ and $3N$ dimensional momentum vector $\bs{p}$ is given by
\begin{align}
    \mu(\bs{x}, \bs{p}) = \underbrace{\frac{1}{Z_{\bs{X}}} \exp\left(-\frac{U(\bs{x})}{\kb T}\right)}_{\probx}\underbrace{\frac{1}{Z_{\bs{P}}} \exp\left(-\frac{|\bs{p}|^2}{2 \kb m T}\right)}_{\mu_{p}(\bs{p})},
\end{align}
where $U(\bs{x})$ is the potential energy function of the atomic system; $m$ is the atomic mass; $T$ is the absolute temperature; $\kb$ is the Boltzmann constant. $Z_{\bs{X}} \equiv \int d\bs{x} \exp\left(-U(\bs{x})/\kb T\right) $ and $Z_{\bs{P}} \equiv \int d\bs{p} \exp\left(-|\bs{p}|^2/2 \kb m T\right)$ are the normalizing constants associated with probability distribution functions $\probx$ and $\mu_{p}(\bs{p})$, and also known as, respectively, the configurational and momentum partition functions.

% expression to compute free energy from the knowledge of the partition function, and discusses the difficulty inherent in computing the partition function.
The main difficulty in computing free energy originates from the fact that the free energy $F$ of a system at $T$ depends on the partition function $Z$ which involves a summation/integration over all possible atomic configurations compatible with the system under investigation, i.e.
\begin{align}
    F = -\kb T \ln \frac{Z_{\bs{X}} Z_{\bs{P}}}{h^{3N}} = -\kb T\ln\left[\frac{1}{h^{3N}} \int d\bs{x} d\bs{p} \exp\left(  -\frac{U(\bs{x}) + \frac{\bs{p}^2}{2m}}{\kb T} \right)\right],
    \label{eq_free_eng_def}
\end{align}
where  $h$ is Planck's constant. The integral is taken over all allowable values of coordinates $\bs{x}$ and momentums $\bs{p}$ belonging to the state. Because of the high dimensionality (number of atoms) of the system, a naive approach of evaluating the integral in Equation~(\ref{eq_free_eng_def}) by enumerating the microstates is highly impractical.

% describes the division of total free energy into momentum and configurational contributions, and presents expressions for relatively difficult-to-compute configurational free energy. 
We can further divide the expression of free energy into configurational and momentum contributions as
\begin{equation}
    \begin{aligned}
        F & = -\kb T\ln \int d\bs{x} \exp\left(  -\frac{U(\bs{x})}{\kb T}  \right)  -\kb T\ln \int d\bs{p} \exp\left(  -\frac{\bs{p}^2}{2m\kb T} \right) + 3N\kb T \ln                      h,                                                                \\
          & =  \underbrace{-\kb T\ln \int d\bs{x} \exp\left(  -\frac{U(\bs{x})}{\kb T}  \right)}_{\text{configurational free energy} (F_{\bs{X}})} \underbrace{-\frac{3N\kb T}{2}\ln 2\pi m\kb T}_{\text{momentum free energy} (F_{\bs{P}})} + 3N\kb T \ln h.
    \end{aligned}
    \label{eq_free_eng_parts}
\end{equation}
where the momentum contribution $(F_{\bs{P}})$ to the free energy is exactly determined by making use of the Gaussian integral. The configurational free energy $F_{\bs{X}}\equiv -\kb T \ln Z_{\bs{X}}$ requires the detailed knowledge of the atomic interactions and the allowed atomic coordinates for the state we are interested in, and is much more difficult to determine. Another expression for the configurational free energy, which will be useful later, is
\begin{align}
    F_{\bs{X}} = \mathbb{E}_{\bs{x}\sim\probx}\left[\underbrace{U(\bs{x})}_{\text{energy contribution}}+\underbrace{\kb T\ln\probx}_{\text{entropy contribution}}\right],
    \label{eq_free_eng_avg}
\end{align}
where $\mathbb{E}_{\bs{x}\sim\probx}[\cdot]$ is the expectation value over random variable $\bs{x}$ with probability distribution function $\probx$. Out of the two contributions (potential energy and entropy) to the free energy, computation of entropy from atomistic simulation is not straightforward and requires resorting to specialized techniques as discussed below. Computation of this configurational free energy is the main focus of this work.

% presents basic working and ingredients of free energy perturbation method, and discusses the critical role played by the reference system used in these type of methods.
The calculation of $F_{\bs{X}}$ through atomistic simulations can be rendered tractable by using the free energy perturbation technique that provides the foundation for widely used equilibrium and non equilibrium free energy methods. We start with a reference state, having a potential energy function $U^0(\bs{x})$, of which the configurational free energy $F^0_{\bs{x}}$ is exactly known
\begin{align}
    F^0_{\bs{X}} =-\kb T \ln Z^0_{\bs{X}} = -\kb T \ln \int d\bs{x}\exp\left(-\frac{U^0(\bs{x})}{\kb T}\right),
    \label{eq_config_free_eng_ref}
\end{align}
where we define configurational partition function for the reference state as $Z^0_{\bs{X}} \equiv \int d\bs{x}\exp(-U^0(\bs{x})/\kb T)$.  We then re-express the configurational free energy of the system of interest from Equation~(\ref{eq_free_eng_parts}) as
\begin{equation}
    \begin{aligned}
        F_{\bs{X}} & = -\kb T \ln \int d\bs{x}\frac{1}{Z^0_{\bs{X}}}\exp\left(-\frac{U(\bs{x})-U^0(\bs{x})}{\kb T}\right)\exp\left(-\frac{U^0(\bs{x})}{\kb T}\right) -\kb T \ln Z^0_{\bs{X}}, \\
                   & = -\kb T \ln \left<\exp\left(-\frac{U(\bs{x})-U^0(\bs{x})}{\kb T}\right)\right>_0 + F^0_{\bs{X}}.
        \label{eq_free_eng_pert}
    \end{aligned}
\end{equation}
Symbol $\langle \cdot \rangle_0$ in the first term of Equation (\ref{eq_free_eng_pert}) indicates the average taken over the microstate samples of the reference system 0. Since the free energy is now expressed as an average over the microstates of the reference system 0, it can be in-principle computed by carrying out atomistic simulation of the reference system. In practice, however, the accuracy and efficiency of the method depends on the closeness or overlap of the two systems in phase space. If there is not much overlap between the two systems, the average estimator will have a large variance and will thus require prohibitively large number of microstate samples for a reliable estimation of the free energy.

% a brief overview of various equilibrium and non equilibrium methods built on the foundation of free energy perturbation method and try to mitigate some limitations intrinsic to the vanilla free energy perturbation method. 
To circumvent this overlapping problem, the widely used equilibrium and non-equilibrium free energy methods envision a sequence of intermediate states between the reference and the system of interest. The free energy difference is inferred from the work done by an effective force driving the switching between the two systems. If the switching is made in an equilibrium fashion, the resulting method is called Thermodynamic Integration (TI)~\cite{Kirkwood1935, Frenkel2001}. TI is nevertheless computationally expensive by the requirement of keeping the intermediate states at thermodynamic equilibrium. Non-equilibrium methods, on the other hand, approach the switching process in a time-dependent fashion~\cite{Watanabe1990}. We start with the equilibrated samples of one system and then slowly transform it into the other system along a non-equilibrium trajectory. For better results, the switching is carried out in both directions: from reference to the system under study and vice-versa. Still, if the reference state does not have significant overlap with the system under investigation, the statistical quality of non-equilibrium methods depends on the rate at which the transition is carried out and the number of trajectories used in the subsequent analysis. For a more comprehensive account of various free energy methods and many associated subtleties, the readers are referred to Refs \cite{Chipot2007, Lelievre2010, Hansen2014}

% A discussion on the issues faced while using manually designed reference system.
From the above discussion, it is evident that the choice of the reference state greatly affects the efficiency, statistical stability, and accuracy of the free energy calculation in the frame work of free energy perturbation method. One widely used reference system in the context of crystalline solids is the Einstein crystal in which all atoms vibrate independently with the same frequency under the influence of a harmonic potential energy function~\cite{DeKoning1996}. However, the Einstein crystal is not the optimum reference state, as the transition from Einstein crystal to the real crystal often involves abrupt changes at the end, and thus a careful attention needs to be paid to the sampling strategy along the switching path in order to reduce statistical error~\cite{DeKoning2005}.

% an assertion that normalizing flow would be able to learn a better reference system.
In the next few sections, we propose a strategy, based on normalizing flow framework, that will be able to learn a reference state to closely resemble the real system. We start with a brief exposition of the normalizing flow itself.

\section{Normalizing flow and free energy as a loss function}
\label{sec_norm_flow}

% describes basic idea of normalizing flow and how it could be relevant to atomistic simulations.

The working principles of normalizing flow based Boltzmann generator was originally proposed by No\'e et al~\cite{Noe2019a}. Here we describe the basic concepts associated with the Boltzmann generator for the sake of completeness, emphasizing the role of the free energy in the construction of the loss function, and the need to handle the motion of the center of mass of the crystal.

Oftentimes, we encounter a situation where it is practically hard to generate samples from a complicated probability distribution. Finite temperature atomistic simulations such as molecular dynamics and Markov Chain Monte Carlo, at equilibrium conditions, also aim to generate atomic configurations from the underlying Boltzmann distribution associated with the canonical ensemble. Normalizing flow is well suited to address such class of problems. The basic idea of normalizing flow is to first generate random samples from an easy-to-sample probability distribution, followed by application of a bijective map to transform the thus generated random samples to the target probability distribution.

% A mathematical description of the various component involved in the working of normalizing flow
To make the idea of normalizing flow more concrete, consider random variable $\bs{z}\in \mathbb{R}^{3N}$, distributed according to probability distribution $\probz$, i.e. $\bs{z} \sim \probz$. We refer to $\probz$ as the base distribution; and in this work it is chosen to be a Gaussian distribution with zero mean vector and identity covariance matrix $(\mathbb{I}_{3N\times 3N})$ from which samples can be generated very easily. Next, by applying a bijective mapping $\bs{f}_{zx}$ on $\bs{z}$, we obtain target random variable $\bs{x} = \bs{f}_{zx}(\bs{z}) \in \mathbb{R}^{3N}$. The corresponding inverse mapping is denoted by $\bs{f}_{xz} = \bs{f}^{-1}_{zx}$ such that $\bs{z} = \bs{f}_{xz}(\bs{x})$. The probability distribution of mapped random variable $\bs{x}$ can be written in terms of the distribution $\probz$ and the Jacobian determinant of the mapping as follows
\begin{align}
    \aprobx = \mu_z\left(f_{xz}(\bs{x})\right) \left| \text{det}\left(\frac{\partial\bs{f}_{xz}(\bs{x})}{\partial \bs{x}}\right) \right| = \mu_z\left(f_{xz}(\bs{x})\right) \gamma_{xz} (\bs{x}),
    \label{eq_aprobx}
\end{align}
where $\gamma_{xz}(\bs{x})$ is the shorthand notation for the  Jacobian determinant of the mapping $\bs{f}_{xz}(\bs{x})$. Because $\gamma_{xz}(\bs{x})$ and $\gamma_{zx}(\bs{z})$ are the Jacobian determinant associated with mapping inverse of each other, we have
\begin{align}
    \gamma_{xz}(\bs{x}) = \frac{1}{\gamma_{zx}(\bs{z})}.
    \label{eq_inv_jac}
\end{align}

\begin{figure}[ht!]
    \centering
    \includegraphics[width=0.8\textwidth]{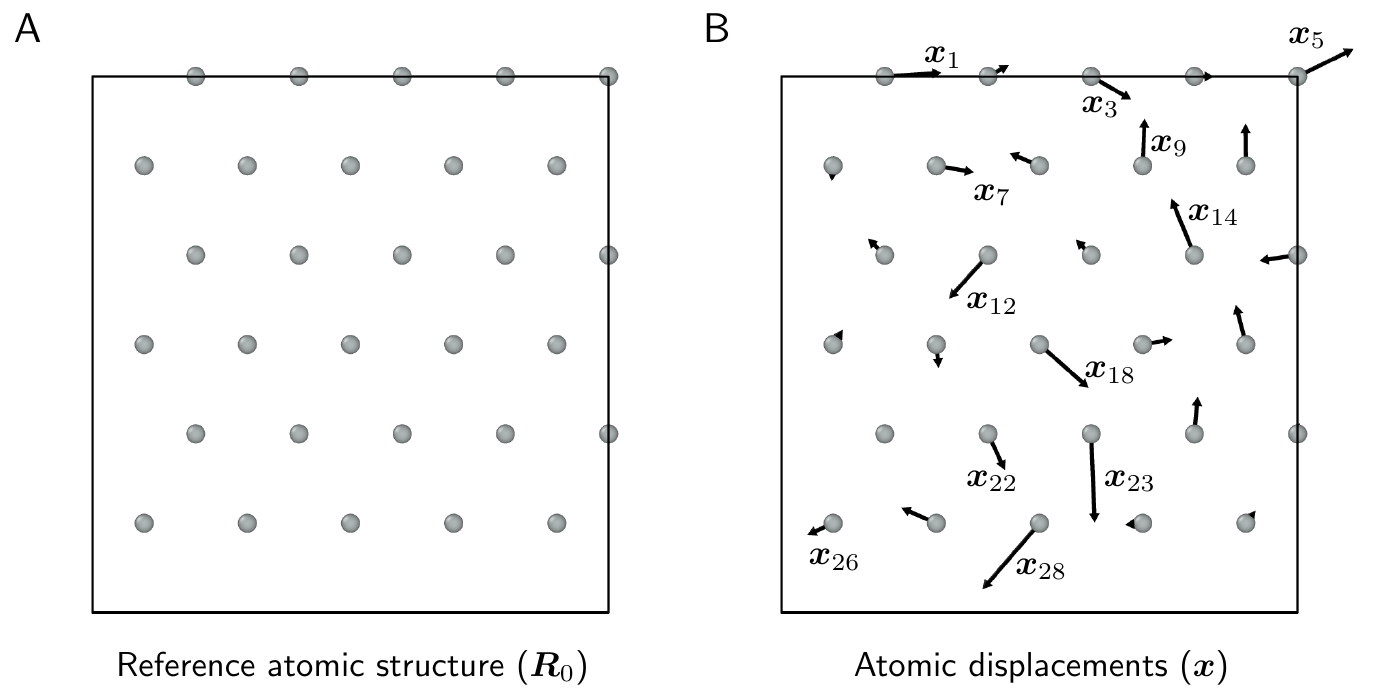}
    \caption{(A) Schematic illustration of the fixed atomic structure $\bs{R}_0$ (grey circles represent atoms) about which displacements $\bs{x}$ are learned. (B) Displacement vector of atom $i$ is depicted by an arrow labelled $\bs{x}_i$.}
    \label{fig_ref_disp}
\end{figure}

% describes connection between abstract mathematical quantities of normalizing flow and an atomic systems in equilibrium. 
In this work, the random variable $\bs{x}$ is interpreted as being the displacement of atoms from a fixed atomic structure $\bs{R}_0$, which are depicted in Figure~\ref{fig_ref_disp}. The potential energy function of the atomic system is represented by $U_0 + U\left(\bs{x}\right)$, where $U_0$ is the potential energy of the fixed atomic structure $\bs{R_0}$. From hereafter, the total potential energy is abbreviated to $U\left(\bs{x}\right)$ to emphasize the dependence of the potential energy on the atomic displacement.

In the case of the canonical $(NVT)$ ensemble, we require that atomic configurations $\bs{x}$ are distributed according to the Boltzmann distribution $\probx$. Thus the task is to learn a mapping $\bs{f}_{zx}(\bs{x})$ such that the transformed distribution $\aprobx$ approach the Boltzmann distribution $\probx$ as closely as possible. To achieve this we introduce parameters $\Theta$ in both of the mapping functions $\bs{f}_{zx}(\bs{z}; \Theta)$ and $\bs{f}_{xz}(\bs{x}; \Theta)$. This work uses real valued non-volume preserving (RealNVP) function as the mapping function $\bs{f}_{zx}(\bs{z}; \Theta)$ of which a specific form is presented in \ref{app_sec_realnvp}. We then need a metric or loss function against which to optimize the parameters $\Theta$.

% Motivates use of free energy as loss function to optimizing parameters involved in normalizing flow, and provides an expression to compute the same.
We first define the variational free energy associated with an arbitrary probability distribution $\aprobx$ of atomic positions $\bs{x}$ with the potential energy $U(\bs{x})$ as follows
\begin{align}
    F[ \aprobx ] = \mathbb{E}_{\bs{x}\sim\aprobx}\left[ U(\bs{x}) + \kb T \ln \aprobx \right].
    \label{eq_var_free_eng}
\end{align}
For the case where $\aprobx$ is the Boltzmann distribution $\probx$, the variational free energy is the same as the Helmholtz free energy as given by Equation~\eqref{eq_free_eng_avg}. Furthermore,  equilibrium statistical mechanical consideration guarantees that the Boltzmann distribution is the unique minimizer of the variational free energy Equation~\eqref{eq_var_free_eng}. Thus, the variational free energy is an appropriate loss function to optimize the parameters $\Theta$ of the mappings, and thus to make distribution $q_x(\bs{x})$ close to the Boltzmann distribution $\probx$. The variational free energy can also be expressed in terms of the base distribution as
\begin{equation}
    \begin{aligned}
        F(\Theta) & = \mathbb{E}_{\bs{z}\sim\probz}\left[ U\left(\bs{f}_{zx}(\bs{z};\Theta)\right) + \kb T \ln \mu_{z}\left(\bs{z}\right) - \kb T \ln \gamma_{zx}(\bs{z}; \Theta) \right].
    \end{aligned}
    \label{eq_free_eng_aprobx}
\end{equation}
We notice that the second term does not depend on the parameters $\Theta$, and thus can be removed to obtain the following loss function that is used in this work as a metric to optimize the mapping
\begin{equation}
    \begin{aligned}
        L\left(\Theta\right) = \mathbb{E}_{\bs{z}\sim\probz} \left[  U\left(\bs{f}_{zx}(\bs{z};\Theta)\right) - \kb T \ln \gamma_{zx}(\bs{z}; \Theta)\right].
    \end{aligned}
    \label{eq_loss_fun}
\end{equation}
The first term in the above loss function tries to reduce the potential energy of and, therefore, narrow the transformed distribution $\aprobx$ towards the lower energy atomic configurations; the second term, however, competes with the first term to encourage the spreading of the distribution $\aprobx$. The strength of the second term depends on the temperature $T$ and acts as a measure of the entropy of the system. We further note that loss function $L$ differs from the variational free energy of the transformed distribution $\aprobx$ only by a constant term which, as evident from Equation \eqref{eq_free_eng_aprobx}, is temperature times entropy of the base distribution. As shown by No\'e et al~\cite{Noe2019a}, the same loss function, Equation~\mref{eq_loss_fun}, can also be obtained by minimizing the Kullback-Leibler divergence between the transformed distribution $\aprobx$ and the target Boltzmann distribution $\probx$.

The translation invariance of potential potential energy function $U(x)$ presents a problem regarding the motion of the center of mass of the atoms. Since potential energy remains unchanged for a rigid body motion, the distribution of atomic displacements $\bs{x}$ optimized using the loss function, Equation~\eqref{eq_loss_fun}, could lead to uncontrolled motion of the center of mass and result in higher entropic contribution to the overall free energy. This could further result in weakly diverging learning regime. To avoid these problems, we propose to augment the potential energy function $U(x)$ by adding a harmonic potential energy term for the center of mass motion. Therefore, we optimize the parameters $\Theta$ under the total potential energy function
\begin{align}
    U_{\rm aug}(\bs{x}) = U(\bs{x}) + \frac{1}{2} k_{\rm com} |\bs{x}_{\rm com}|^2,
    \label{eq_aug_pot_eng}
\end{align}
where $k_{\rm com}$ is the stiffness of the harmonic spring attached to the center of mass, and $\bs{x}_{\rm com}$ is the displacement vector of the center of mass. The additional contribution to the total free energy (including momentum term) arising due to the harmonic motion of the center of mass has the following analytical form
\begin{align}
    F_{\rm com} = -3 \kb T \ln \left(\frac{\kb T}{\hbar} \sqrt{\frac{Nm}{k_{\rm com}}}\right),
    \label{eq_free_eng_com}
\end{align}
which is subtracted from the overall free energy to obtain the final free energy as described in the next section.

\section{Learned distribution as a reference state for free energy computation}
\label{sec_reweighting}

% interpretation of the base distribution as an atomic system in equilibrium under the canonical ensemble.
As discussed in the preceding sections, normalizing flow produces a distribution of atomic configurations $\aprobx$ by transforming a simple base probability distribution $\probz$. The variables $\bs{z}$ and $\bs{x}$ are understood as denoting the atomic displacements from a fixed atomic structure $\bs{R}_0$ with potential energy $U_0$. We can interpret these two distributions as two thermodynamic systems in the canonical ensemble with their respective potential energy functions. We consider that the base probability distribution $\probz$ defines thermodynamic system I (TSI) with associated potential energy function $U^{\rm I}(\bs{z})$, i.e.
\begin{align}
    \probz = \frac{1}{Z_{\rm I}} \exp\left( -\frac{U^{\rm I}(\bs{z})}{\kb T} \right).
    \label{eq_probz_eng}
\end{align}
We further assume that due to the simplicity of the base distribution, its partition function $Z_{\rm I}$ is exactly known and, in return, its free energy $F_{\rm I} \equiv -\kb T \ln Z_I$. For instance, if every $3N$ components of $\bs{z}$ are normally distributed with zero mean and unit standard deviation, as used in this work, the corresponding potential energy function is $U^{\rm I} (\bs{z}) = \bs{z}^2/2$, and partition function takes the value $Z_{\rm I} = (2\pi\kb T)^{3N/2}$.

% interpretation of the transformed distribution as an atomic system in equilibrium under the canonical ensemble.
The transformed distribution obtained from the application of normalizing flow on the base distribution is termed thermodynamic system II (TSII). To obtain the potential energy function of this TSII, we make use of Equations \mref{eq_aprobx} and \mref{eq_probz_eng} to write
\begin{equation}
    \begin{aligned}
        \aprobx                           & = \frac{1}{Z_I} \exp\left( -\frac{U^{\rm II}(\bs{x}) }{\kb T} \right),       \\
        \text{where }  U^{\rm II}(\bs{x}) & =  U^{\rm I}\left(\bs{f}_{xz}(\bs{x})\right) - \kb T\ln \gamma_{xz}(\bs{x}).
    \end{aligned}
    \label{eq_eng_TSII}
\end{equation}
It is clear from the above equation that under the canonical ensemble, the potential energy function of TSII is $U^{\rm II}(\bs{x})$. Under this potential energy function free energy of TSII is the same as that of TSI, i.e. $F^{\rm II} = F^{\rm I} \equiv -\kb T\ln Z_{\rm I}$, which is known by design.

% discussion on the suitability of the transformed distribution as the reference system for free energy methods.   
We note that the bijective mapping function $\bs{f}_{zx}$ is learned such that TSII should approach ever more closely the target Boltzmann distribution $\probx$ with the augmented potential energy function $U_{\rm aug}(\bs{x})$. This further implies that the potential energy of TSII, $U^{\rm II}(\bs{x})$, approximates the augmented potential energy function $U_{\rm aug}(\bs{x})$ except for an additive constant, i.e $U^{\rm II}(\bs{x}) \approx U_{\rm aug}(\bs{ x }) + {\rm constant}$. This property is shown in Figure~\ref{fig_real_learned_eng}, where the learned energy $U^{\rm II}(\bs{x})$ of the generated configurations obtained from a trained normalizing flow has a strong correlation with the `real' augmented potential energy $U_{\rm aug}(\bs{x})$.

However, due to limitations in the expressive power of the mapping function (e.g. the limited number of parameters) and computational resources (limiting the training time), the transformed distribution $\aprobx$ would not be exactly the same as the target distribution $\probx$, although they should be close. This closeness, nevertheless, is well suited to be exploited in the framework of free energy perturbation methods, as discussed in Section \ref{sec_free_eng_pert}, to compute free energy of the system of interest.

% Provides a concrete mathematical expression to determine unbiased free energy using normalizing flow framework.
Since, as shown in the preceding paragraphs, the potential energy and free energy of TSII are fully accessible, it can be used as a reference system to compute free energy of the target system with the augmented potential energy function $U_{\rm aug}(\bs{x})$ given by Equation~\eqref{eq_aug_pot_eng}. Using Equation \mref{eq_free_eng_pert}, configurational free energy of the target system can be determined as
\begin{equation}
    \begin{aligned}
        F_{\bs{X}, \text{aug}} & = -\kb T \ln \mathbb{E}_{\bs{x}\sim \aprobx}\left[ \exp\left( -\frac{U_{\rm aug}(\bs{x}) - U^{\rm II}(\bs{x})}{\kb T} \right) \right] + F^{\rm II},                                                                            \\
                               & = -\kb T \ln \mathbb{E}_{\bs{z}\sim \probz}\left[ \exp\left( -\frac{U_{\rm aug}\left(\bs{f}_{zx}(\bs{z})\right) - U^{\rm I} \left(\bs{z}\right) - \kb T \ln \gamma_{zx}(\bs{z})}{\kb T} \right) \right] - \kb T \ln Z_{\rm I},
    \end{aligned}
    \label{eq_free_eng_reweight}
\end{equation}

% a brief summary of the section, and a statement on the similarity between free energy perturbation method and reweighting by importance sampling.
Thus, Equation \mref{eq_free_eng_reweight} offers a way to compute the exact unbiased free energy of the system. Estimation of the average required in the first term of Equation \mref{eq_free_eng_reweight} over the base distribution can be performed efficiently by generating multiple samples from the distribution  $\probz$ in parallel. We further remark that exactly the same Equation \mref{eq_free_eng_reweight} can also be obtained through the route of importance sampling which is frequently-used reweighting method in statistics~\cite{Noe2019a}.

After determining the configurational part of the free energy from Equation~\eqref{eq_free_eng_reweight}, we add the momentum contribution Equation~\eqref{eq_free_eng_parts} and subtract the center of mass contribution Equation~\eqref{eq_free_eng_com} to obtain the final unbiased total free energy
\begin{align}
    F = F_{\bs{X}, \text{aug}} + F_{\bs{P}} - F_{\rm com}
    \label{eq_free_eng_final}
\end{align}

\begin{figure}[ht!]
    \centering
    \includegraphics[width = \textwidth]{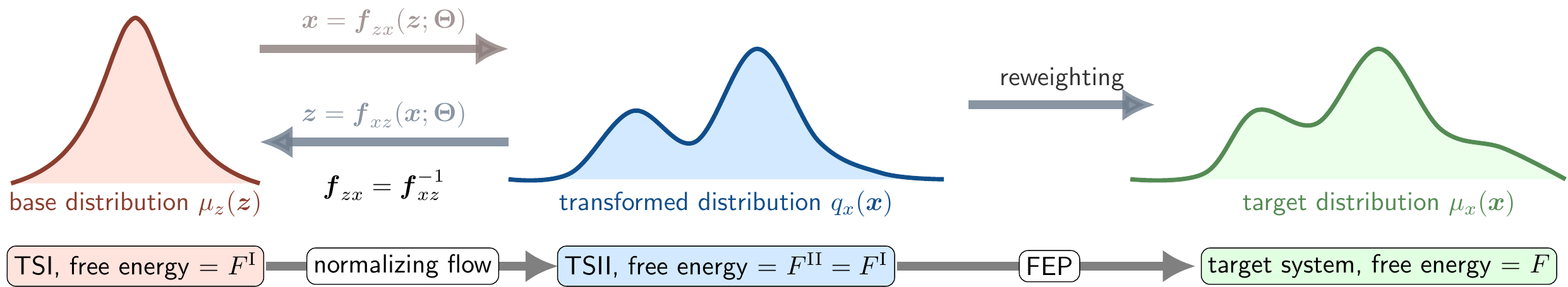}
    \caption{Schematic illustration of various steps involved in free energy the free energy of the system of interest using the normalizing flow. The simple base distribution representing TSI is transformed into a distribution representing TSII. TSII has the same free energy as TSI, and the associated potential energy function of TSII is close to that of the target system. A reweighting procedure based on the free energy perturbation (FEP) method is finally applied to TSII to determine the free energy of the target system.}
    \label{fig_schematic_flow}
\end{figure}

The workflow of the normalizing flow from base distribution to computing the free energy of the target system is laid out schematically in Figure~\ref{fig_schematic_flow}.

\section{Application to perfect crystal and vacancy formation in diamond-cubic Si}
\label{sec_application_si}

% a description of the atomic system (diamond cubic Si) on which normalizing flow framework is applied.
In this section we apply the normalizing flow framework, as presented in the preceding sections, to determine the temperature-dependent free energy associated with the real system of interest. We first compute the Gibbs free energy at zero stress condition, which is equal to the Helmholtz free energy, of perfect diamond-cubic Si crystal. %We choose this system due to the availability of carefully computed results in the literature using other well-established methods that provide a high-fidelity benchmark for the normalizing flow framework.

We then turn our attention to vacancy, a crystalline defect, which is fundamental in controlling materials transport and determining several mechanical properties. Vacancies play a prominent role at high temperatures by diffusing in the material in the field of other crystalline defect such as grain boundaries and dislocations. Thus quantifying the thermodynamic properties of vacancy is necessary to build an understanding of its wide-ranging influence on material properties. We particularly focus on the formation free energy which is an important thermodynamic quantity  governing the equilibrium population of vacancy at finite temperatures.

\subsection{Computational details}

% description of the simulation box, conditions,  and interatomic potential employed
A periodic simulation box is constructed by replicating the diamond-cubic unit cell (each containing 8 atoms with 0 K  lattice parameter 5.430 \AA) three times in all three directions. The simulation box thus contain 216 atoms. For vacancy study, the vacancy is introduced by removing one atom from the simulation box of perfect crystal and minimize the energy keeping the simulation box dimensions fixed. Interatomic interactions of Si atoms are described by using Stillinger–Weber (SW) potential of which parameters are listed in Ref.~\cite{Stillinger1985}. To compute the Gibbs free energy at finite temperatures, we first compute the equilibrium lattice parameters at corresponding temperatures by running isobaric $NPT$ simulations at zero stress conditions and averaging the simulation box sizes. The simulation boxes constructed with these finite temperature lattice parameters are the fixed atomic structures $\bs{R_0}$ used in the normalizing flow framework about which the displacements $\bs{x}$ are learned. The Helmholtz free energy computed at the fixed volume of zero stress is the same as Gibbs free energy at zero stress.

% description of the various parameters involved in normalizing flow that are used in this work.
We now describe the structure and various components involved in RealNVP based normalizing flow used in this work. The base distribution for perfect crystal $\probz$ is a multivariate Gaussian over 648 (3$\times$216) dimensional random vector $\bs{z}$ with zero mean and identity covariance matrix, i.e. no correlation exists among its components. For vacancy, the base distribution is defined over 645-dimensional (3 $\times$ 215) space. Successive transformations, as described in \ref{app_sec_realnvp}, are applied on $\bs{z}$ sampled from $\probz$ to obtain the displacement vector $\bs{x}$. To apply the coupling layer of RealNVP, we divide the vector into two halves, each having 324 components. In each coupling layer, scaling function $\bs{s}(\cdot;\Theta)$ and translation function $\bs{t}(\cdot;\Theta)$ are two neural networks whose hyperparameters are listed in Table~\ref{table_hyperparameter}. In total we apply ten coupling layers, each with its own separate set of parameters in scaling and translation functions. We finally multiply thus transformed vector with an optimizable scalar parameters to control the initial magnitude of the displacement vectors $\bs{x}$. This step of scaling is necessary to prevent the unreasonably large displacement that can cause energy to become too high and lead to instability in the optimization process.

\begin{table}
    \centering
    \caption{ Hyperparameters of the two neural networks: scaling $\bs{s}$ and translation $\bs{t}$.(see \ref{app_sec_realnvp} for details) These hyperparameters are the same across 30 Coupling layers used in the mapping function. The optimizable parameters, however, are tuned independently in all the coupling layers during training, i.e. we don't us parameter sharing among the neural networks.}
    \begin{tabular}{l| p{2.5 cm} p{2.5 cm}}
        \toprule
        neural network          & $\bs{s}$() & $\bs{t}()$ \\
        \midrule
        input dimension         & 324        & 324        \\
        output dimension        & 324        & 324        \\
        number of hidden layers & 3          & 3          \\
        hidden layer dimension  & 500        & 500        \\
        non linearity           & tanh       & LeakyReLU  \\
        \bottomrule
    \end{tabular}
    \label{table_hyperparameter}
\end{table}

% description of the training settings.
During the training of the normalizing flow, 128 samples from the base distribution $\probz$ are generated to modify parameters based on the gradient of the loss function of Equation \mref{eq_loss_fun}. The training is performed using ADAM algorithm with multiple learning rates: $10^{-3}$ for the first 4000 steps, $10^{-4}$ for the next 4000 steps and $10^{-5}$ for the last 2000 steps. After training the normalizing flow, the total free energy is determined using Equation~\eqref{eq_free_eng_final} where average is calculated using 10000 samples drawn from the base distribution.

% libraries, codes used in this work for energy computation and deep learning parts.
The computation of energies of atomic systems are performed by using open source MD code LAMMPS~\cite{Plimpton1995}; and the normalizing flow, including neural networks, is constructed and trained using the python library Pytorch~\cite{Steiner2019}.

\subsection{Results}

This section presents the results obtained from the application of the normalizing flow to perfect crystal and vacancy formation in diamond-cubic Si.

\subsubsection{Gibbs free energy of perfect diamond-cubic crystal}

\begin{figure}[ht!]
    \centering
    \includegraphics[width = 0.5\textwidth]{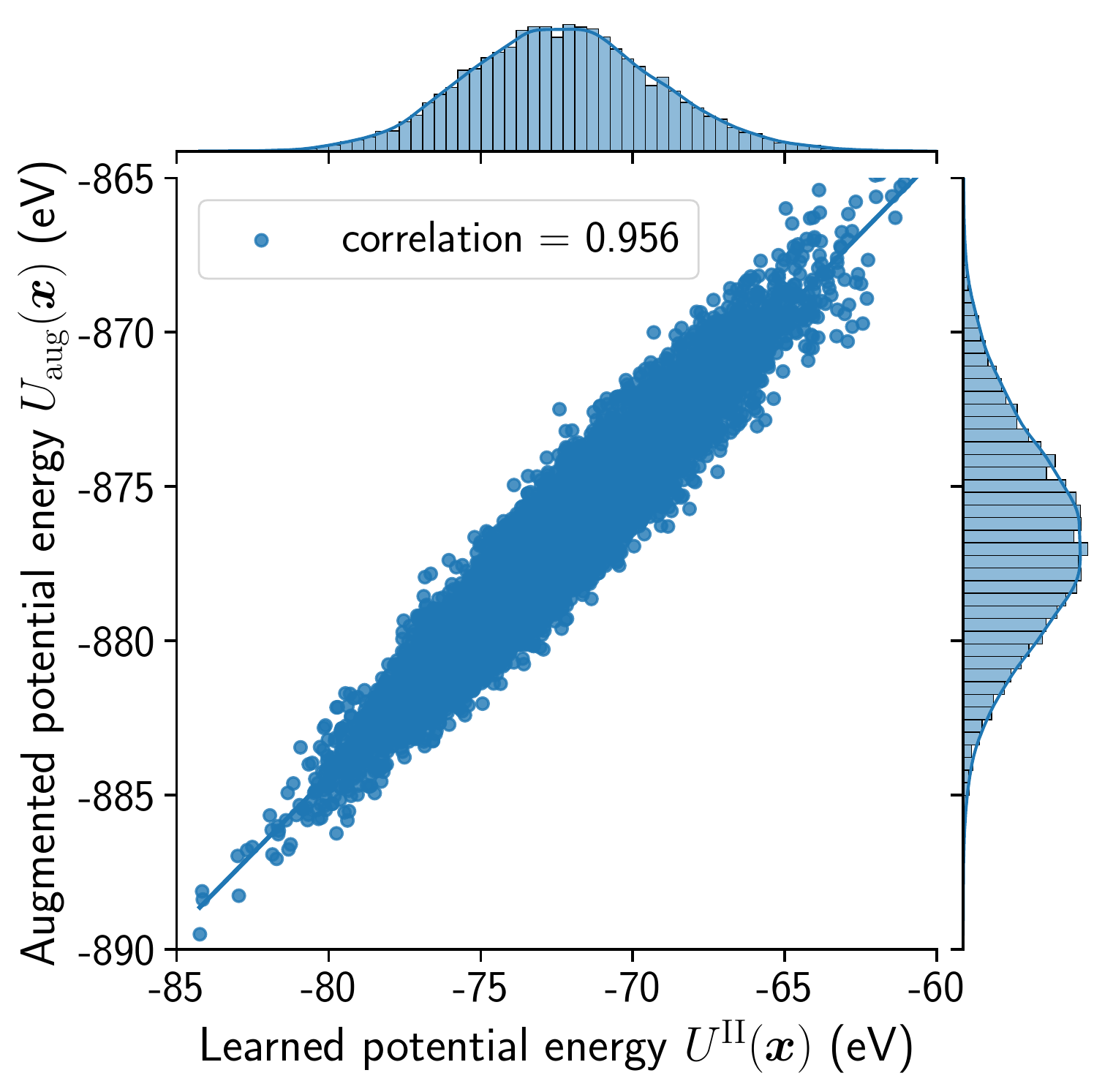}
    \caption{Joint distribution of the augmented potential energy function $U_{\rm aug}(\bs{x})$ and the learned potential energy $U^{\rm II}(\bs{x})$ of TSII. The atomic configurations $\bs{x}$ are generated from the normalizing flow trained at 2000 K. The potential energy $U^{\rm II}(\bs{x})$ of of TS II has strong correlation with the augmented potential energy $U_{\rm aug}(\bs{x})$ that includes energy from the SW potential and energy due to harmonic center of mass potential. }
    \label{fig_real_learned_eng}
\end{figure}

Figure~\ref{fig_real_learned_eng} shows the comparison between the learned potential energy $U^{\rm II}(\bs{x})$ of TSII, Equation~\eqref{eq_eng_TSII}, and the augmented potential energy function $U_{\rm aug}(\bs{ x })$, Equation~\eqref{eq_aug_pot_eng}, for the atomic configurations $\bs{x}$ generated from the normalizing flow trained at 2000 K. The plot of the joint distribution shows a strong correlation between the two energy functions. Thus the normalizing flow is able to learn a thermodynamic system TSII with the potential energy function $U^{\rm II}(\bs{ x })$ close to the augmented potential energy function $U_{\rm aug}(\bs{x})$.

% a description of what is computed and presented as results.
We now present the Gibbs free energy of Si computed from normalizing flow framework, and compare with the values computed from the adiabatic switching method following Ref.~\cite{Ryu2008}. The Gibbs free energy at a particular temperature is equal to Helmholtz free energy determined using the equilibrium lattice constant at the corresponding temperature. Total Free energy is given in Equation \mref{eq_free_eng_parts}, in which the configurational free energy contribution is evaluated using Equation \mref{eq_free_eng_reweight}.

\begin{figure}[ht!]
    \includegraphics[width=\textwidth]{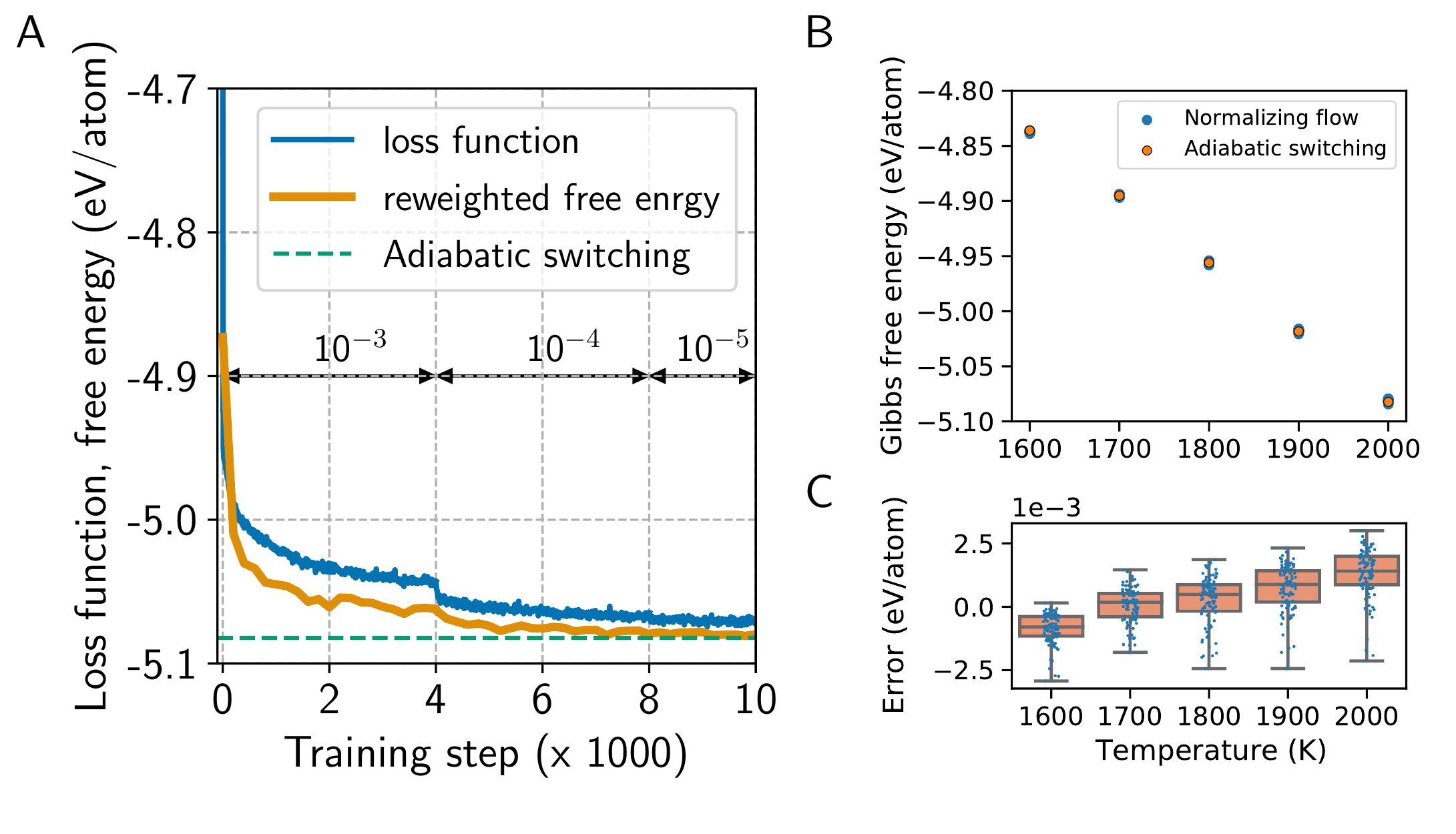}
    \caption[]{(A) Variation in loss function, including momentum contribution and entropy of base distribution, with training steps at 2000 K temperature is plotted with blue line. Multiple learning rates($10^{-3}, 10^{-4}$ and $10^{-5}$) used during different stages of training are shown.  Change in the Gibbs free energy, obtained by the perturbation procedure, with training steps is also plotted with orange color. The free energy calculations using the importance sampling reweighting are performed after every 200 training steps. The green dashed horizontal line shows the free energy obtained from the adiabatic switching method. (B) Blue circles show a scatter plot of 100 difference calculations of the Gibbs free energy per atom using the normalizing flow framework as a function of temperature. The Gibbs free energy values as computed from the adiabatic switching method are shown with orange circles for comparison. (C) Plot of the distribution of the  difference between normalizing flow and adiabatic switching values for 100 different values obtained fom the normalizing flow.}
    \label{fig_loss_free_eng}
\end{figure}

% describes the variation in loss function, reweighted configurational free energy as training proceeds
Figure \ref{fig_loss_free_eng}(A) presents the variation in the loss function as a function of training steps. It is clear that the loss function decreases during the training. At each training step, 128 independent random samples are generated to compute the correction in parameters according to the gradient signal obtained from the loss function. Thus, the whole learning framework is that of unsupervised learning, and requires no prior data set. We also show in Figure~\ref{fig_loss_free_eng} (A) the variation in the total free energy, computed using Equation \mref{eq_free_eng_reweight} as training proceeds. This free energy is computed after every 200 steps by averaging over 1000 samples generated from the normalizing flow. The free energy converges near the value obtained from the adiabatic switching method. The free energy values are presented in Table~\ref{table_free_eng_perfect_crystal} for both normalizing flow and adiabatic switching methods.

% presents Gibbs free energy of diamond-cubic Si as a function of temperature and comparison with  available literature values.  
Figure \ref{fig_loss_free_eng}(B) shows the total Gibbs free energy, including momentum contribution as depicted in Equation \mref{eq_free_eng_parts}, of a perfect diamond cubic Si crystal at temperatures ranging from 1600 to 2000 K. Since we are using finite number of samples (10000), the free energy computed from Equation~\eqref{eq_free_eng_final} is itself stochastic. Therefore, we compute the free energy for 100 different sets of 10000 random samples generated from the trained normalizing flow. We also show the Gibbs free energy of the same system  by using the adiabatic switching method~\cite{Ryu2008}. Since the scatter in the normalizing values is too small (of the order of $10^{-3}$ eV/atom) to be visible at the scale of the figure, we show the distribution of the difference between 100 normalizing flow values and adiabatic switching value in Figure~\ref{fig_loss_free_eng}(C). The result from adiabatic switching is just from one switching simulation (for $10^6$ MD steps) and we didn't attempt to quantify its error. However, both results show good agreement with each other establishing that normalizing flow can be employed to determine free energies accurately and efficiently.

\begin{table}
    \centering
    \caption{ Gibbs free energy per atom (eV/atom) as a function of temperature of perfect diamnond-cubic Si crystal as computed from the normalizing flow and the adiabatic switching methods. We list only the average values of the free energies obtained from both methods.}
    \begin{tabular}{l| p{2 cm} p{2 cm} p{2 cm} p{2 cm} p{2 cm}}
        \toprule
        Temperature (K)     & 1600    & 1700    & 1800    & 1900    & 2000    \\
        \midrule
        Normalizing flow    & -4.8369 & -4.8951 & -4.9556 & -5.0175 & -5.0809 \\
        Adiabatic switching & -4.8360 & -4.8951 & -4.9559 & -5.0183 & -5.0822 \\
        \bottomrule
    \end{tabular}
    \label{table_free_eng_perfect_crystal}
\end{table}

We now compare the computational costs of the two free energy methods as measured in both wall-clock time and number of potential function evaluations. The training of the normalizing flow takes around 6.5 hours while computing the free energy using Equation~\eqref{eq_free_eng_final} (reweighting is performed with 10000 samples) using the trained normalizing flow takes around 2 minutes. On the other hand, the adiabatic switching method takes around 45 minutes to perform one pair of forward and backward switching between harmonic and real crystals. Thus, despite a little longer training process, the normalizing can be more efficient in collecting a large statistics from the trained normalizing flow. We, however, want to emphasize that we did not invest much effort in optimizing the training process and shortening the training time. Furthermore, one million evaluations of potential energy and atomic forces were performed during the normalizing flow training step; and one adiabatic switching trajectory also involves one million molecular dynamics step. The fact that we need a smaller number of samples (10000 in this work) to compute reweighted free energy using the trained normalizing flow as opposed to the one million MD steps of the adiabatic switching further emphasizes the efficiency of the normalizing flow framework.

%For example, training only requires the evaluation of potential energy but not the atomic forces, which is required in MD simulations for adiabatic switching.  However, we didn't take advantage of this and used the same LAMMPS function that computes both potential energy and forces during training. (Is this true?)

\subsubsection{Formation free energy of monovacancy}

To compute the formation free energy of monovacancy at finite temperatures, we train two normalizing flows: one for the perfect crystal with and another for the crystal containing a vacancy. After training the two normalizing flows, the free energies of the two systems are computed using Equation~\eqref{eq_free_eng_final}. The formation free energy of monovacancy $F_{\rm f}^{\rm vac}$ is then computed as
\begin{align}
    F_{\rm f}^{\rm vac} = F^{\rm vac} - \frac{N-1}{N} F^{\rm per},
\end{align}
where $F^{\rm vac}$ is the free energy of the crystal containing a vacancy, $F^{\rm per}$ is the free energy of the perfect crystal, and $N$ is the number of atoms in the prefect crystal ($N = 216$ in this work).

\begin{figure}[ht!]
    \centering
    \includegraphics[width = 0.5\textwidth]{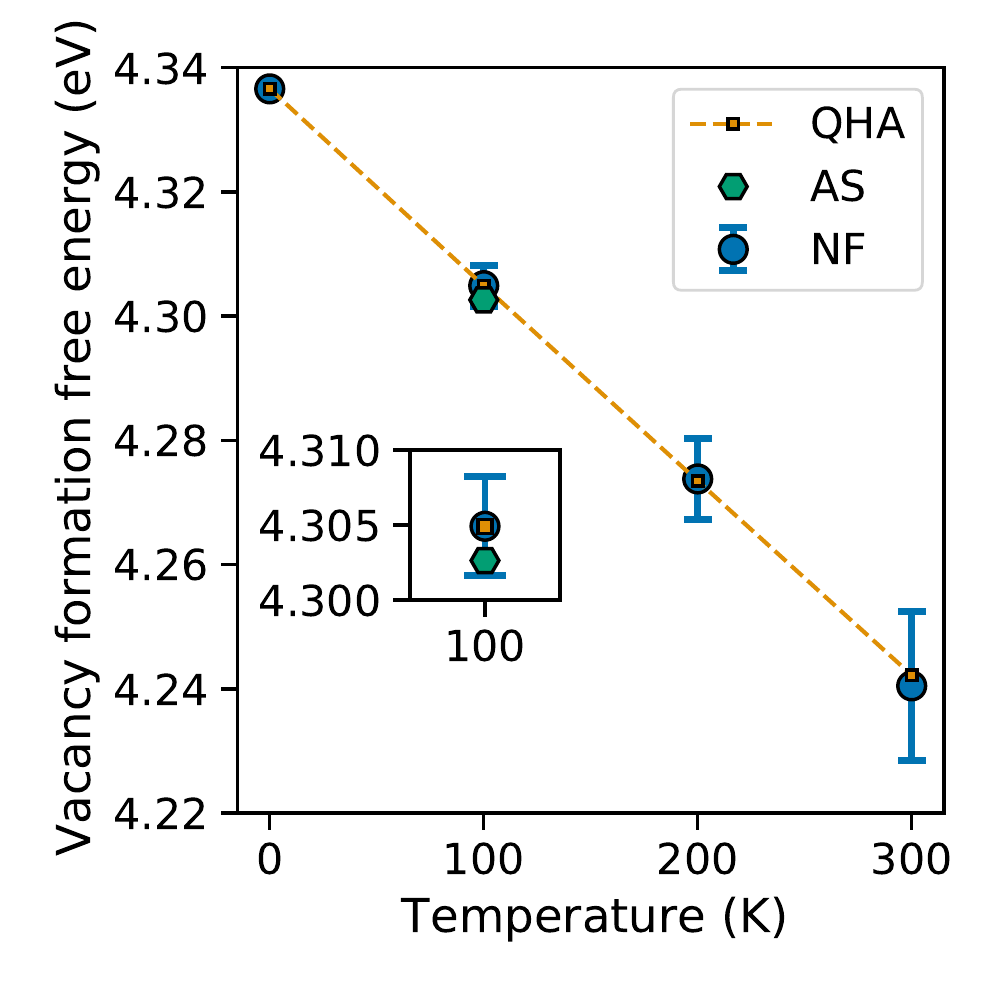}
    \caption{Temperature dependent formation free energy of monovacancy in diamond cubic Si crystal at zero stress condition. Free energies are computed from three independent methods: quasi harmonic approximation (QHA), adiabatic switching (AS) (at $100$~K only), and normalizing flow (NF). The error-bar for the normalizing flow calculations show the standard deviation for 100 different reweighting calculations, each involving 10000 samples. The inset shows the zoomed-in plot of different values of the vacancy formation free energy at 100 K  as obtained from the three different methods.}
    \label{fig_vac_form_free_eng}
\end{figure}

Figure~\ref{fig_vac_form_free_eng} shows the formation free energy of monovacancy at temperatures ranging from 0 to 300 K in diamond-cubic Si as determined from three independent methods: normalizing flow (NF), quasi harmonic approximation (QHA)~\cite{Ramirez2012a, Xie1999}, and adiabatic switching method. The reasons for focusing onto the relatively lower temperatures are two-fold. First is to avoid the complication arising from the vacancy diffusion~\cite{Foiles1994, Cheng2018}. Indeed, we found that at temperatuers above $100$~K, the vacancy jumps during the adiabatic switching simulation, producing very large error bars for the free-energy estimate.  This makes our current implementation of the adiabatic switching not applicable to vacancy formation free energy calculation above $100$~K. Second, the quasi harmonic approximation becomes reasonably accurate at lower temperatures and thus provides a good reference value for the comparison.

\begin{table}
    \centering
    \caption{ Variation of the formation free energy (eV) of monovacancy in diamnond-cubic Si crystal at zero stress condition with temperature as determioned from the normalizing flow and the quasi harmonic approximation methods. The adiabatic switching method is able to provide the vacancy formation free energy only at 100 K temprature due to the observation of the vacancy diffusion at higher temperatures. The value at 0 K is obtained by minimizing the potential energy. The values shown in the table correspond to the average free energy of monovacancy formation determined from the three methods.}
    \begin{tabular}{l| p{2 cm} p{2 cm} p{2 cm} p{2 cm} }
        \toprule
        Temperature (K)              & 0      & 100    & 200    & 300    \\
        \midrule
        Normalizing flow             & 4.3366 & 4.3049 & 4.2738 & 4.2405 \\
        Quasi harmonic approximation & 4.3366 & 4.3049 & 4.2735 & 4.2422 \\
        Adiabatic switching          & 4.3366 & 4.3026 & -      & -      \\
        \bottomrule
    \end{tabular}
    \label{table_formation_free_eng_vac}
\end{table}

From Figure~\ref{fig_vac_form_free_eng}, it is manifest that the three methods employed to compute the free energy agree reasonably well establishing the capacity of the normalizing flow to be used in computing the temperature dependent free energy of crystalline defects. From the slope of the vacancy formation energy as a function of temperature, we obtain the formation entropy of the vacancy as 3.72 $\kb$. The formation free energies of monovacancy at temratures ranging from 0 to 300 K are listed in Table~\ref{table_formation_free_eng_vac}.

\section{Conclusion and discussion}
\label{sec_conc_disc}

% A  brief summary of the main result of this work.
In this work, we demonstrated that normalizing flow based framework can be employed to compute reasonably accurate free energy for a solid. Specifically, we computed the zero stress Gibbs free energy of diamond cubic Si at temperatures from 1600 to 2000 K, and showed that there is a good agreement with the available literature values.

% summary of the vacancy formation free energy step
We further show that the normalizing flow can be used to determine the formation free energy of monovacancy defect in diamond-cubic crystal at temperatures ranging from 0 to 300 K. This is particularly encouraging due to the fact that material properties emerges from the finite-temperature behavior of various crystalline defects and their mutual interactions, i.e. vacancies, interstitials, solutes, dislocations, stacking fault etc. The challenges involved with computing the free energy associated with defects often leads to resorting to the approximation of the defect free energy by potential energy or tractable local harmonic approximation~\cite{Ahmad2018, Ahmad2019, Ahmad2020, Wu2018, LeSar1989, Ryu2011}. The accuracy of such approximation becomes questionable at high temperatures where the anharmonic nature of underlying potential energy function begins to show appreciable effect~\cite{Cheng2018}. Thus the normalizing flow framework is promising in efficient quantification of the finite-temperature mechanisms associated with various crystalline defects, and, in turn, aiding in understanding, controlling and engineering technologically important materials.

% discussion on the vacancy formation free energy calculation
Computing the vacancy formation energy at high temperatures is challenging due to issues arising from the diffusion process~\cite{Cheng2018}. As an example in case, we observe the diffusion of vacancy during the adiabatic switching process even at temperatures as low as 200 K. We do not encounter the vacancy diffusion in the normalizing flow framework till 300 K temperature. This is presumably caused by the setup of the problem in which we learn distribution of atomic displacements about a fixed metastable reference state instead of evolving the whole system iteratively as done in MD and MC simulations. Furthermore, the additional constraining of the center of mass by introducing a harmonic potential may also penalize any atom undergoing a systematic large displacement over and above the random thermal fluctuations. However, the framework as presented in this work does not guarantee to prevent the vacancy diffusion entirely at arbitrarily temperatures, and we, therefore, must practice care while investigating defect having multiple metastable states separated by relatively low energy barrier.

% discussion on the computational costs of the two methods
A comparison of the time taken by the normalizing flow and the adiabatic switching shows that while the training of the normalizing flow takes a relatively long time, the normalizing can still be much more efficient in collecting a large statistics to obtain reliable results. We, further note that the training time of the normalizing flow can further be improved by optimizing the various steps involved in the training. For example, we use the CPU serial version of LAMMPS in this work to compute energy and forces of the generated atomic configurations. This entails the transfer of data between GPU and CPU during each energy and force computation step. The GPU version of LAMMPS could be leveraged to accelerate the training process. Moreover, various hyperparameters  involved in the normalizing flow architecture - learning rate, number of coupling layers, number of hidden dimensions and layers - can be optimized to speed up the training and inference step enhancing the efficiency of the normalizing flow even further.

% discussion on the ability of the normalizing flow to produce a system with known free energy
The normalizing flow has been shown to transform a simple base distribution into a distribution close to the target Boltzmann distribution. We show that the normalizing flow produces a thermodynamic system with known free energy and its associated potential energy function is close to the thermodynamic system under investigation. This enables the use of the learned thermodynamic system as a reference state to determine the free energy of the system of interest.

% discussion on the use of other reweighting procedure
The reweighting procedure, necessitated by the difference between the transformed (reference) and the target distribution, was performed using the importance sampling or the free energy perturbation method. This is the simplest possible reweighting which still provides accurate results. Since we have access to the normalized probability density of the transformed distribution, more elaborate procedures could be used for the reweighting, for example adiabatic switching~\cite{Watanabe1990} and annealed importance sampling~\cite{Neal2001}.

% Interpretation of the workflow presented in this work as targeted free energy perturbation method.
The normalizing flow along with reweighting by importance sampling can also be understood as the targeted free energy perturbation method as described in Ref.~\cite{Jarzynski2002}. Basically, the transformed distribution is a learned reference system rather than designed a priori. This obviates the work involved in computing various properties to be included in the reference system. For instance we would need to have access to the normal mode frequencies of the target system to setup a harmonic approximation-crystal as the reference system. Thus normalizing flow provides a workflow to obtain an optimized reference system in a principled manner for use in the subsequent free energy perturbation methods.

% advantage of the unsupervised setting of the normalizing flow. 
Training of the normalizing flow proceeds in a unsupervised manner. Data to train the network are generated from the base distribution, and thus are unlimited. Each training step encounters new set of randomly sampled data. All these features reduces the possibility of overfitting of the neural networks.

% Potential advantages of using transformed distribution as proposal in Metropolis-Hastings algorithm to sample unbiased equilibrium atomic configurations. 
Furthermore, normalizing flows provide a way to efficiently generate equilibrium samples from the Boltzmann distribution using the Metropolis-Hastings algorithm~\cite{Nicoli2020}. In this scheme, transformed distribution can be used to generate proposals at each step. As the transformed distribution is close to the target, the acceptance probability of the proposed configurations would be much higher, and the generated samples will have small correlation with good mixing. Moreover, the proposal can be generated in parallel accelerating the process significantly.

% possible future work to compare other possible mapping functions. 
In this work, we used RealNVP as the invertible mapping function. RealNVP has advantage in that its operations are quite simple which leads to efficient learning regime. However, RealNVP has some  restriction in its expressivity, necessitating composition of several function units, resulting in a large number of parameters. Other more complex and universal mapping functions have been proposed in the literature. However, a systematic study into their  memory footprints versus ease of training still needs to be undertaken in the context of free energy computation of atomic systems.

% discusses the need to modify the existing normalizing framework before applying to extended crystal defects involving millions of atoms. 
Modeling of extended crystalline defects, such as dislocation, grain boundaries, and their interactions can require thousands to millions of atoms~\cite{Zepeda-Ruiz2017, Bertin2020}. Thus, application of normalizing flow based algorithms to the extended crystalline solid system with defect may suffer from the scalability of the existing normalizing flow algorithms: the number of parameters in the deep neural network to be learned scales with square of the number of atoms. Thus, atomic modeling of the crystalline solids through the normalizing flow algorithms requires further modifications and will be addressed in future work.

% possible impact of the present work.
This work has established that normalizing flow provides an efficient way to determine accurate free energy of solid systems. We believe that this normalizing flow framework would help in gaining a deeper understanding of various finite temperature mechanisms underlying wide range of crucial behavior of crystalline solids, and in turn facilitate the design of new class of technologically important materials.

\section*{Acknowledgements}
RA acknowledges the financial support of this work through a research grant from the Swiss National Science Foundation entitled ``Investigation into finite temperature atomic-scale crystal plasticity through generative deep learning'' (project \# P2ELP2\_199806). %The authors would like to thank --- for critically reading and providing insightful comments on earlier versions of the manuscript.

\appendix
\setcounter{figure}{0}
\setcounter{table}{0}
\renewcommand{\theequation}{\Alph{section}.\arabic{equation}}

\section{Form of the mapping function: real valued non volume preserving (RealNVP) mapping}
\label{app_sec_realnvp}

% Desired properties of the mapping function to be used in normalizing flow.
Here, we describe the form of the mapping function $\bs{f}_{zx}(\bs{z})$ itself which is the critical component of normalizing flow. The form of the mapping function should be: invertible, which is intrinsic to the basic working of normalizing flow; expressive, i.e. ability to approximate arbitrarily complex probability distribution function; efficient in computing the Jacobian determinant, i.e. determining the determinant of the jacobian of the mapping function should not entail heavy computational resources and time which is essential for tractable and feasible learning/optimization regime of parameters $\Theta$.

\begin{figure}[ht!]
    \includegraphics[width=\textwidth]{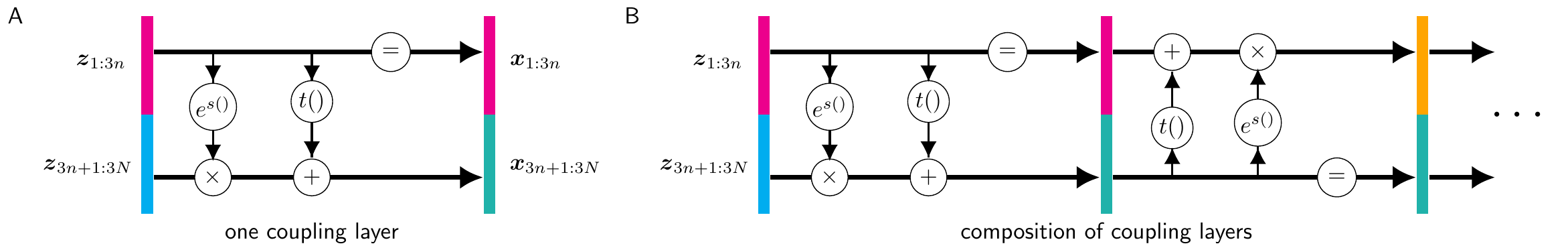}
    \caption[]{(A) A schematic depiction of the operation of single coupling layer that transform $\bs{z}$ into $\bs{x}$. Note that one half (magenta colored) of the $\bs{z}$ vector is kept unchanged, and only the other half (cyan colored) is transformed. (B) To increase the expressivity of RealNVP, multiple coupling layers are composed in sequence. Note that the unchanged half is alternated to transform the whole vector after two operations of coupling layer. $\bs{s}()$ and $\bs{t}$() functions appearing in coupling layers are modeled by two neural networks, and each coupling layer can have its own independent parameters.}
    \label{fig_relanvp}
\end{figure}

% Mathematical description of the mapping function (RealNVP) used in this work.
Several mapping functions have been proposed in the literature satisfying the above three properties in varying degrees~\cite{Kobyzev2020,Papamakarios2019}. In this work we use the so-called real valued non volume-preserving (RealNVP) mapping, proposed by Dinh et al~\cite{Dinh2019} to transform the base Gaussian distribution. which is also used in Ref.~\cite{Noe2019a} The transform of $\bs{z}$ variable in RealNVP takes place as follows
\begin{equation}
    \begin{aligned}
        \bs{x}_{1:3n}    & = \bs{z}_{1:3n},                                                                                                                       \\
        \bs{x}_{3n+1:3N} & = \bs{z}_{3n+1:3N} \odot \exp\left( \bs{s} \left( \bs{z}_{1:3n}; \Theta \right)  \right) + \bs{t}\left( \bs{z}_{1:3n}; \Theta \right),
    \end{aligned}
    \label{eq_realnvp_one}
\end{equation}
where $\odot$ is notation for Hadamard or element-wise product;  $\bs{s}:\mathbb{R}^{3n}\rightarrow\mathbb{R}^{3N-3n}$ and $\bs{t}:\mathbb{R}^{3n}\rightarrow\mathbb{R}^{3N-3n}$  are, respectively, scaling and translation functions that contain all the parameters of the mapping function. There is no restriction on these two functions and can possess any level of complexity. In this work these rwo functions are modeled as two deep neural networks. The operation of Equation \mref{eq_realnvp_one} is schematically depicted in Figure \ref{fig_relanvp}(A). It can be readily checked that this mapping is invertible and inverse can be computed quite easily. The Jacobian matrix of the transformation is given by
\begin{align}
    \frac{\partial \bs{x}}{\partial \bs{z}} = \begin{bmatrix}
        \begin{array}{c|c}
            \mathbb{I}_{3n\times 3n}                                 & \mathbb{O}_{3n\times 3N-3n}                                    \\[1em]
            \hline                                                                                                                    \\
            \frac{\partial \bs{x}_{3n+1:3N}}{\partial \bs{z}_{1:3n}} & \text{diag}\left(\exp\left(\bs{s}(\bs{z}_{1:3n})\right)\right)
        \end{array}
    \end{bmatrix},
\end{align}
where $\mathbb{I}_{3N\times 3N}$ is an identity matrix, and $\mathbb{O}_{3n\times 3N-3n}$ is a zero matrix. Since the Jacobian matrix is a triangular matrix, its determinant is simply the product of the diagonal terms, which is
\begin{equation}
    \begin{aligned}
        \gamma_{zx}(\bs{z})     & = \left|\det \left(\frac{\partial \bs{x}}{\partial \bs{z}}\right) \right| = \prod_{i = 1}^{3N-3n}\exp\left(s_i(\bs{z}_{1:3n})\right), \\
        \ln \gamma_{zx}(\bs{z}) & = \sum_{i=i}^{3N-3n}s_i\left(\bs{z}_{1:3n}\right).
    \end{aligned}
\end{equation}
Therefore, the Jacobian determinant of the mapping can be efficiently determined without any additional expensive computation.

% A discussion on the disadvantage of using RealNVP and introduces composition of functions to mitigate it to some extent 
One limitation of the mapping function presented in Equation \mref{eq_realnvp_one} is that one part of the vector $\bs{z}$ is untransformed which clearly imposes severe restriction on the class of distribution function that can be obtained from this mapping. To resolve this expressivity issue to some extent, we can compose many such functions in sequence. One function unit being composed is called coupling layer in the normalizing flow literature. The part of the vector that is kept unchanged is alternated between subsequent coupling layers as shown in Figure \ref{fig_relanvp}(B). Expressivity of the overall mapping can be increased by applying many coupling layers. Furthermore, parameters of all the coupling layers need not be different and can be shared among compatible layers.

% Computation of Jacobian and inverse of the composition of functions.
The invertibility of the overall mapping is ensured by the fact that all the coupling layers being composed are invertible. Computing inverse itself is easy because inverse of a composition of function is composition of inverse of the functions in reverse order. The Jacobian determinant of the overall mapping is computed by using the fact that determinant of the Jacobian of composition of functions is the product of the Jacobian of the constituent functions involved in the composition. To write these facts in mathematical form, suppose $\bs{x}$ is obtained by applying  $p$ number of functions $f_1, f_2, f_3, \dots, f_p$ is in sequence, i.e. $\bs{x} = f_p \circ f_{p-1} \circ \dots \circ f_1 (\bs{z})$. The inverse and the Jacobian determinant of the mapping is expressed as
\begin{equation}
    \begin{aligned}
        \bs{z}      & = f^{-1}_1 \circ f^{-1}_2 \circ \dots \circ f^{-1}_p (\bs{x}), \\
        \gamma_{zx} & = \prod_{i=1}^{p} \gamma_{i},
    \end{aligned}
\end{equation}
where $\gamma_i$ is the Jacobian determinant of the function $f_i$ evaluated at its input.

% weighs pros and cons of RealNVP and points to other possible mapping functions proposed in the literature.
RealNVP is quite efficient in performing computations as it involves only a few simple mathematical operations. However this efficiency comes at the cost of reduced expressivity. To approximate a complex probability distribution functions, a large number of coupling layers may be required. For a detailed analysis of the expressivity of RealNVP and other mapping commonly used in normalizing flow, the readers are encouraged to consult Refs.~\cite{Kobyzev2020,Papamakarios2019}.

\addcontentsline{toc}{section}{References}
\bibliographystyle{unsrt}

\end{document}